\documentclass[aps,prl,twocolumn,showpacs,showkeys,floatfix]{revtex4}

\usepackage{graphicx}
\usepackage{dcolumn}
\usepackage{bm}

\newcommand{\pdiffl}[2]{\frac{\partial #1}{\partial #2}}

\newcommand{\dfrac}[2]{\displaystyle\frac{#1}{#2}}

\begin{document}


\title{Improved analysis of converging shock radiographs}

\date{August 16, 2019, updates to March 10, 2022
  -- LLNL-JRNL-832834}

\author{Damian C. Swift}
\email{dswift@llnl.gov}
\author{Andrea L. Kritcher}
\author{Amy Lazicki}
\author{Natalie Kostinski}
\author{Brian R. Maddox}
\author{Madison E. Martin}
\author{Tilo D\"oppner}
\author{Joseph Nilsen}
\author{Heather D. Whitley}
\affiliation{%
   Lawrence Livermore National Laboratory,
   7000 East Avenue, Livermore, California 94550, USA
}

\begin{abstract}
We previously reported an experimental platform to induce
a spherically-convergent shock in a sample using laser-driven ablation,
probed with time-resolved x-ray radiography,
and an analysis method to deduce states along the principal shock Hugoniot
simultaneously with the x-ray opacity.
We have now developed a modified method of analysis that is numerically
better-conditioned and faster, and usually provides a better representation
of the radiograph with correspondingly lower uncertainties.
The previous approach was based on optimizing parameters in a model
of the density distribution as a function of radius and time,
warped to follow loci such as the shock and the outside of the sample.
The converging shock configuration can be described more efficiently in terms
of the shocked density and sound speed, 
expressed as functions of the shock speed
Studies of the Hugoniot from various theoretical equations of state (EOS)
indicate that, in the typical range of states explored by these experiments,
these functions can be described by low-order polynomials.
Similarly, few-parameter functions were found suitable for representing
the variation of x-ray opacity with shock pressure.
This approach was found to perform better in most cases than
an alternative method based on parameterization of the EOS.
\end{abstract}

\pacs{07.35.+k, 47.40.Na, 64.30.-t}
\keywords{shock physics, equations of state, laser ablation, radiography, carbon}

\maketitle

\section{Introduction}
Simultaneous compression and heating of matter by a shock wave 
has been used for over seven decades to induce and study the highest 
pressures accessible in bulk matter \cite{shock}.
We have developed an experimental platform using pulsed lasers to
generate a spherically-converging shock
\cite{Kritcher2014,Doeppner2018,Swift2018,Kritcher2020,Swift2021}.
The speed and compression of the shock are inferred from x-ray radiography.
Because these are simultaneous measurements of two parameters of the shock
in the sample material itself, the measurement is absolute as opposed to
being with respect to a reference material.
Even if driven by a constant ablation pressure, the shock becomes
stronger as it propagates radially inward, because of isentropic compression
in the flow behind which causes a progressive rise in shock pressure.
Thus each experiment can sample a range of shock pressures, at the cost of
a more complicated analysis.

Because, along the x-ray trajectories forming the radiograph,
matter immediately behind the shock is obscured by matter in other states,
the profile of mass density must be reconstructed in order to 
determine the shock state.
Although this reconstruction can in principle be performed in a deterministic
way by working radially inward by the process of Abel inversion \cite{Abel1826},
this approach accumulates noise. We have found it advantageous to make
more direct use of the known state ahead of the shock in an iterative
analysis in which the profile is represented by analytic functions 
of radius and time,
whose parameters are adjusted until the inferred radiograph best matches the
observation.
In consequence, an obvious concern is the degree to which the choice of
function imposes bias or structure on the model and the inferred shock states.
We investigated this sensitivity by performing analyses with different
choices of function, including tabulations, and by analyzing the radiograph
by dividing it into subsets of different length in the temporal domain,
and found that the variation in shock Hugoniot from the choice of
functional form was several times smaller than the uncertainty from noise
in the radiograph and from systematic uncertainties in the time and space
scales.
However, we were not able to deduce the shock Hugoniot to the precision
expected from our original assessments of the experimental uncertainties,
suggesting inaccuracy or redundancy in the underlying structure of the
radiographic model.
Furthermore, given more recent experimental campaigns in which we collected
data from more than one shot on the same type of sample,
we started to consider how we might infer the best-fitting shock Hugoniot
to measurements from multiple experiments, for which the
space-time object model is inherently unsuitable.

A different strategy would be to parameterize the equation of state (EOS),
and perform simulations of each converging shock experiment, adjusting the
parameters until the simulated radiograph matches the measured record.
We originally disfavored this approach because it relies on additional
numerical parameters whose value is not closely constrained, such as the
treatment of the shock discontinuity (usually via an arbitrary artificial
viscosity) and the resolution in the radial direction.
More fundamentally, current capabilities to predict ablative loading
are not sufficiently accurate, and therefore the drive would also have to be
expressed in parameterized form and adjusted simultaneously with the EOS.
This additional degree of freedom makes
iterative optimization over a hydrocode model computationally much slower
and more expensive than the approach via an object model.

Here we report different approaches to constructing object models for
analyzing radiographic shock data, showing example data from experiments
on carbon (initially solid density in the diamond structure).

\section{Experimental configuration}
\begin{figure}
\begin{center}
\includegraphics[scale=0.30]{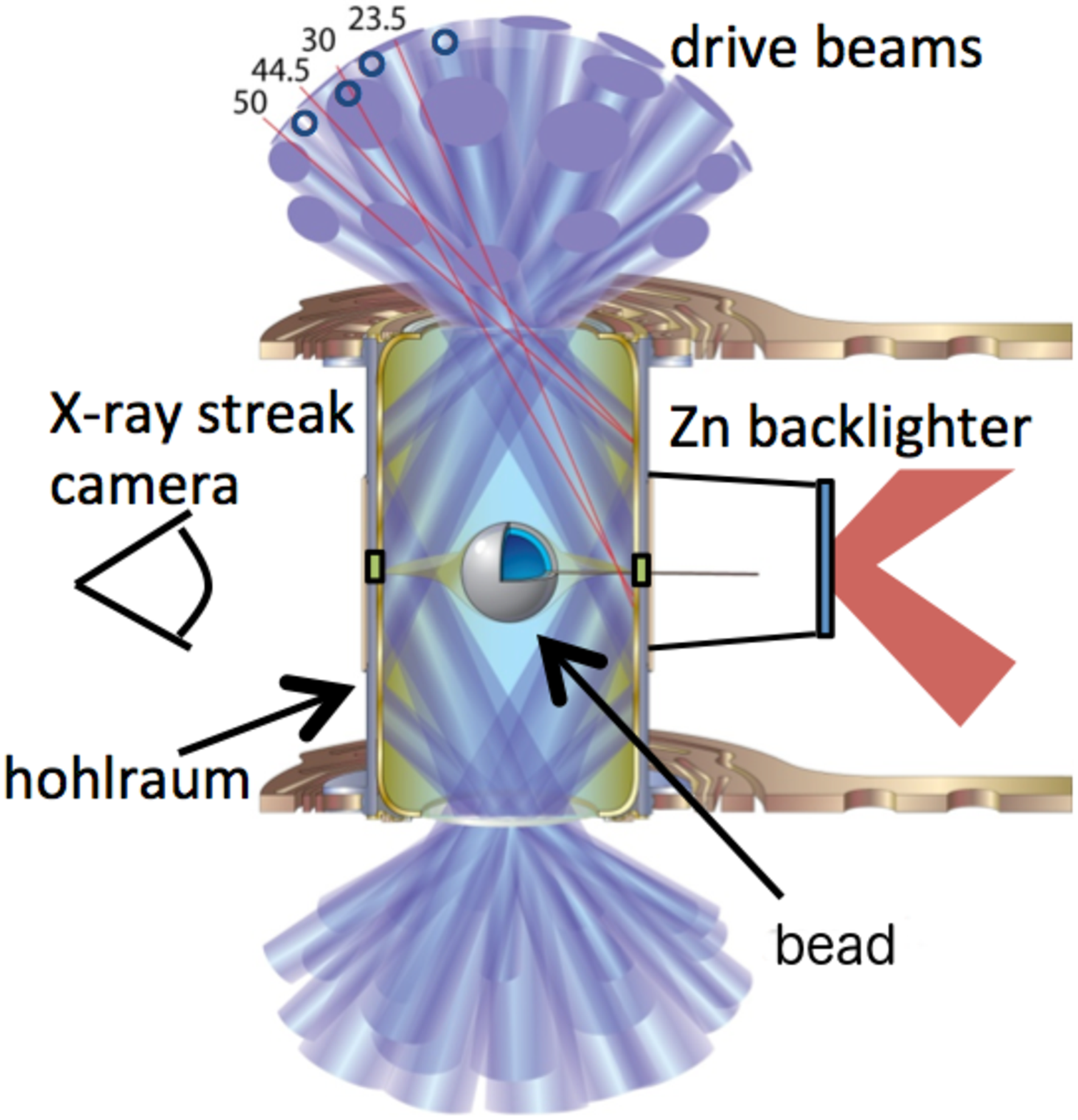}
\includegraphics[scale=0.20]{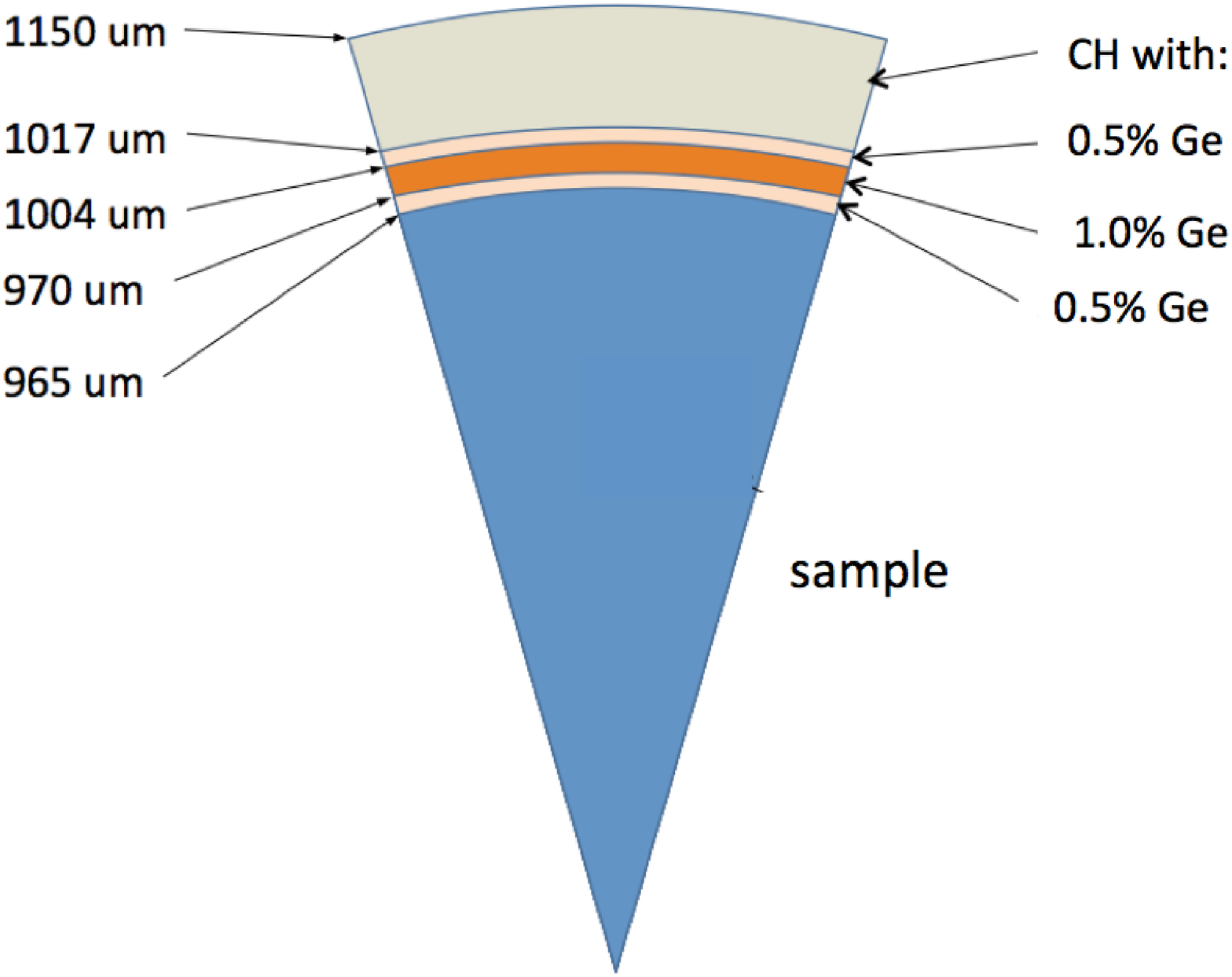}
\end{center}
\caption{Schematic of hohlraum-driven converging-shock experiment.
   Wedge diagram shows sequence of shells comprising spherical target bead.}
\label{fig:exptschem}
\end{figure}

The experiments were performed at the National Ignition Facility (NIF).
The experimental configuration was as described previously
\cite{Kritcher2014,Swift2018,Doeppner2018},
and is summarized here for convenience.
The sample material was prepared in the form of a sphere.
A uniform layer of a polymer was deposited over the surface, to act as
an ablator during the experiment.
The coated sphere was mounted within a Au hohlraum \cite{hohlraum},
30\,$\mu$m thick, 5.75\,mm diameter and 9.42\,mm high,
with a gas fill of 0.03\,mg/cm$^3$ He to impede filling of the hohlraum by ablated Au.
168 beams of the NIF laser were used to heat the hohlraum.
The resulting soft x-ray field within the hohlraum induced a pressure pulse
in the ablator,
driving a shock into the sample.
The overall configuration and laser pulses were based on ICF designs,
to take advantage of synergies in fabrication and also the large development
effort performed to give uniform drive conditions over the surface of the
bead \cite{icf}.
(Fig.~\ref{fig:exptschem}.)

We consider data from two experiments, N161016-3 and N140529-2, in which the
laser drive was 300 and 800\,kJ respectively.
The latter was based on the `high foot' ICF drive \cite{icf,Hopkins2015};
the former was the foot of the drive continued for 5\,ns.
The temperature history of soft x-rays in the hohlraum was calculated using
radiation hydrodynamics and measured by the {\sc Dante} filtered diode system
\cite{Dewald2004}.
The peak temperature was around 205\,eV for the low drive and 275\,eV for the high.

\begin{figure}
\begin{center}\includegraphics[scale=0.65]{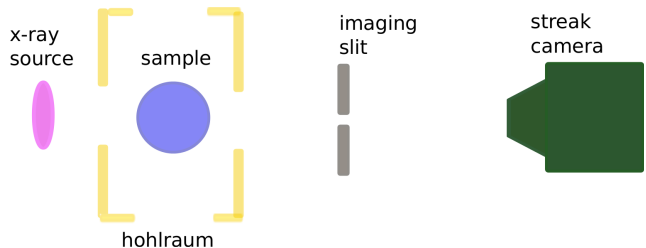}\end{center}
\caption{Radiograpic configuration (not to scale).}
\label{fig:rgconfig}
\end{figure}

The ablation-induced shock wave strengthened as it propagated
toward the center of the sample.
X-ray radiography was used to measure the variation of attenuation
across the diameter of the sample, from which the shock trajectory,
mass distribution and opacity in the sample could be deduced as described
below.
The x-ray source was a Zn foil, heated by eight beams to produce a plasma
that emitted strong He-like radiation (i.e. from atoms stripped of all but
two electrons).
Slits were cut in the hohlraum wall to enable transmission of the x-rays
through the sample; diamond wedges were mounted in the slits to impede
their closure by ablated Au.
The transmitted x-rays were imaged through a slit in a Ta foil onto 
an x-ray streak camera (Fig.~\ref{fig:rgconfig}).

The streak radiograph was used to reconstruct the radial distribution
of mass density, as a function of time.
As described previously \cite{Swift2018}, the presence of undisturbed material
ahead of the shock provided a strong constraint on the inference of
the change in attenuation across the shock front.
In order to take advantage of this constraint, the analysis was performed by
adjusting a parameterized representation of the distribution of mass density
until the corresponding simulated radiograph matched the measured radiograph.

In shot N140529-2, the shock became strong enough that the opacity of shocked
material to 9\,keV x-rays decreased significantly.
The outer edge of the diamond sphere was clearly visible on the radiograph,
so the variation in apparent mass from the profile-matching fit could be used
to infer the change in opacity as the shock passed, as was demonstrated
previously for polystyrene \cite{Swift2021}.

\begin{figure}
\begin{center}\includegraphics[width=0.5\textwidth]{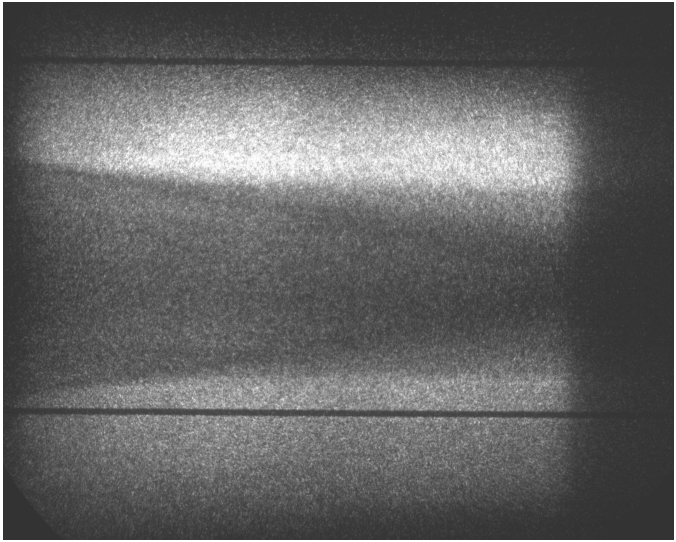}\end{center}
\caption{X-ray streak radiograph, NIF shot N161016-3 (low drive).}
\label{fig:rg161016}
\end{figure}

\begin{figure}
\begin{center}\includegraphics[width=0.5\textwidth]{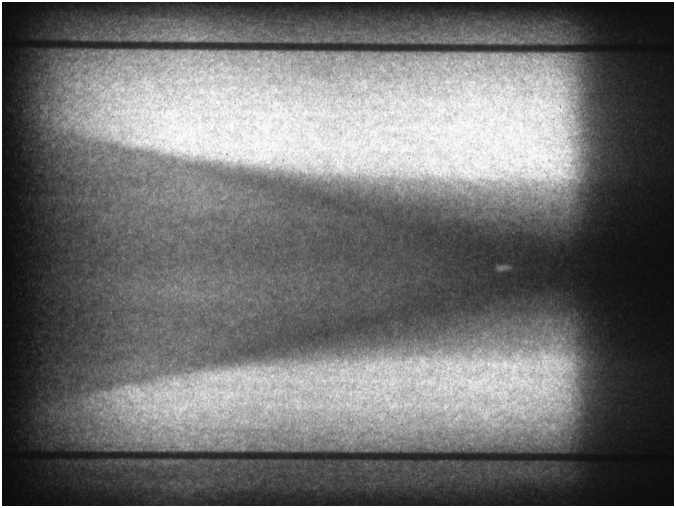}\end{center}
\caption{X-ray streak radiograph, NIF shot N140529-2 (high drive).}
\label{fig:rg140529}
\end{figure}

\section{Analysis}
As discussed previously \cite{Swift2018},
the reconstructed radius-time distribution of mass density $\rho(r,t)$ gives 
an absolute measurement of the shock Hugoniot over a range of pressures,
from the position of the shock $r_s(t)$ and hence its speed $u_s(t)$,
and the mass density immediately behind the shock,
$\rho_s(t)=\rho(r_s(t),t)$.
Simultaneous knowledge of $u_s$ and $\rho_s$ gives the complete mechanical
state behind the shock by solving the Rankine-Hugoniot relations \cite{shock}
representing the conservation of mass, momentum, and energy across the shock,
\begin{eqnarray}
u_s^2 & = & v_0^2\dfrac{p-p_0}{v_0-v} \\
u_p & = & \sqrt{\left[(p-p_0)(v_0-v)\right]} \\
e & = & e_0 + \frac 12 (p+p_0)(v_0-v)
\end{eqnarray}
where $v$ is the reciprocal of the mass density $\rho$,
$e$ is the specific internal energy, $p$ the pressure,\footnote{%
For materials in which material strength is significant, $p$ is the
normal stress rather than the mean pressure.
}
and subscript `0' denotes material ahead of the shock (with $u_p=0$).
The state ahead of the shock is known, leaving five quantities to be
determined ($v, p, e, u_p, u_s$).
If any two of these quantities are measured, the Rankine-Hugoniot equations
determine the rest.
In particular,
\begin{equation}
p = p_0 + \dfrac{u_s^2}{v_0^2}\left(v_0-v\right).
\end{equation}
Thus the mechanical state on the Hugoniot can be deduced directly from
the distribution of mass density, without reference to any other material
used as a standard, as is the case in some other experimental configurations,
and so the measurement is absolute.

We previously reported analyses using a direct parameterization of 
the density distribution $\rho(r,t)$, and the opacity either assumed constant
or inferred along the Hugoniot by the variation in apparent mass enclosed 
within a Lagrangian marker \cite{Swift2018,Swift2021}.
One complication is that it is not immediately clear what
functional forms to use to represent $\rho(r,t)$, and so it was necessary
to choose forms with a high degree of generality and thus a large number
of parameters, and to investigate the sensitivity of the deduced Hugoniot
to the forms used. The analysis thus became relatively ill-conditioned and
computationally intensive, and we did not achieve the precision expected
from our original assessments of uncertainty.
The opposite extreme would be to perform an analysis by parameterizing
models in forward, hydrocode, simulations of the experiment.
We considered this approach undesirable 
partly because of potential concerns about
the accuracy of the simulations, for example from the use of artificial
viscosity to stabilize shocks by smearing them spatially and temporally,
but mostly because our ability to predict the drive from the fundamental
experimental input of the laser power history is not adequately accurate,
and therefore the drive itself would also require parameterization,
introducing an additional family of parameters and computational expense.

In the present work, we improve the analysis in essence by considering 
a hierarchy of models with different degrees of physical constraint.
The previous approach assumed only that mass was conserved and that the
state ahead of the shock was known, with further implicit constraints
(intended to be as few as possible and motivated by the goodness of fit) 
from the choice of $\rho(r,t)$ functions.
Here we investigate the effect of assuming in addition that the experiments
probe matter with a unique shock Hugoniot, but no further hydrodynamic
constraints from the large-scale flow.
A hydrocode-based analysis is equivalent to appling further constraints
on the nature of the flow behind the shock.

An important addition to the analysis presented here is the use of the density
gradient immediately behind the shock, which can be related to its acceleration.
This is a key aspect of converging shock experiments.
The acceleration of the shock depends on
the spatial gradient of density behind the shock $\partial\rho/\partial r$,
the variation of pressure with density variation along isentrope
   $\partial p/\partial\rho|_s$,
the rate at which characteristics catch up with shock (its sonicity)
   $c(u_s)+u_p(u_s)-u_s$,
and the variation of shock speed with pressure $\partial u_s/\partial p_s$.
The acceleration of the shock can be expressed with respect to 
density gradient, the Hugoniot and its derivative, and the sound speed
on the Hugoniot:
\begin{equation}
\dot u_s = \pdiffl \rho r \left.\pdiffl p\rho\right|_s
   \left[c(u_s)+u_p(u_s)-u_s\right] \pdiffl{u_s}{p_s}
\label{eq:acc}
\end{equation}
where $\partial p/\partial\rho|_s=c^2$.
Given an analytic EOS, the sound speed can be calculated.
The use of a functional fit for $r_s(t)$ gives 
analytic expressions for $u_s(t)$ and $\dot u_s(t)$.
We have generaally found that $r_s(t)$ can be expressed as a power law,
also giving power laws for $u_s(t)$ and $\dot u_s(t)$.

This connection between the density gradient and the acceleration can be
regarded as employing some of the apparent self-similarity of converging
shocks, in local and differential form as opposed to seeking a global 
solution.
In particular, the analysis can take advantage of the 
local consistency of acceleration with speed.

\begin{figure}
\begin{center}\includegraphics[width=0.5\textwidth]{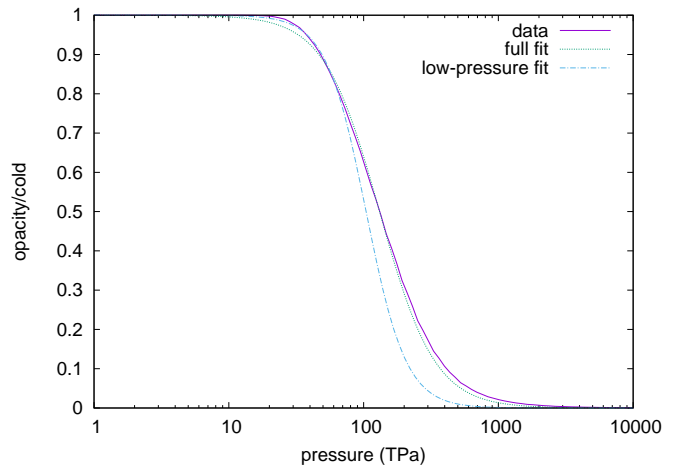}\end{center}
\caption{Example model opacity, compared with Fermi function fits.}
\label{fig:opacmodel}
\end{figure}

Another important change was the use of an
algebraic function to describe the x-ray opacity along the shock Hugoniot.
We investigated the use of several functions, but eventually settled on
an expression derived from the Fermi occupation function,
which captured the essential behavior of an initially constant opacity
followed by a smooth decrease.
Temperature was not determined in these experiments:
the Hugoniot relations allow mechanical quantities to be determined, 
including the internal energy, but not the temperature.
However, $p\sim T$ in the regime where $k$-shell ionization occurs,
and we found it possible represent the variation in theoretical
opacity with a Fermi-like function in pressure,
\begin{equation}
\mu=\dfrac{\mu_0}{1+\exp\left[(\log_{10}p-\log_{10}p_c)/w\right]}.
\end{equation}
This function was not able to reproduce the variation of theoretical
opacity accurately over the full range of its decrease,
but captured it to better than 1\%\ for the variation from
cold to around a 50\%\ decrease.
Low-order polynomials, fitted to the theoretical opacity,
were used to represent the subsequent variation $\partial\mu/\partial\rho|_s$
along isentropes from states along the Hugoniot.
(Fig.~\ref{fig:opacmodel})

With a model-based opacity as opposed to an unfold, 
the mass constraint used to break the redundancy in seeking to deduce
density as well as opacity becomes implicit instead of explicit.
The use of the model provides a constraint of uniqueness and also makes the
opacity much less sensitive to noise in the radiograph.

\subsection{Parameterize density field $\rho(r,t)$}
In the approach applied previously \cite{Swift2018,Doeppner2018}, 
the mass density was parameterized directly
as a function of radius and time, $\rho(r,t)$.
The radial variation was defined in terms of a scaled radius between the 
time-varying shock and marker loci.
As discussed above, the density model is inherently specific to each experiment,
does not naturally admit guidance from theoretical models,
and does not lend itself to deducing Hugoniot data from multiple experiments
simultaneously.

The approach chosen previously
\cite{Kritcher2020,Swift2021}
to account for the varying opacity was, at each instant of time, 
to integrate inward from the marker layer to the shock, and infer the
change in opacity from the apparent mass enclosed.
As with Abel inversion, this unfolding technique accumulates error.
It would be possible to apply the model-based technique described above
with a parameterized $\rho(r,t)$ field, but we did not explore this 
additional permutation of approaches.

\subsection{Parameterize natural Hugoniot functions}
The inherent Hugoniot measurement from the radiographic, converging shock
experiments is the mass density as a function of the shock speed,
$\rho(u_s)$.
In the density field analysis above, the shock trajectory $r_s(t)$ was
represented independently of the density along the shock locus,
$\rho[r_s(t),t]$, and neither are inherent properties of the EOS.
$\rho_s(u_s)$ is however a property of the EOS (from the Hugoniot, which is
unique for a given initial state.)
Represented in this form, it is readily possible to
analyze multiple experiments simultaneously.

\begin{figure}
\begin{center}\includegraphics[scale=0.72]{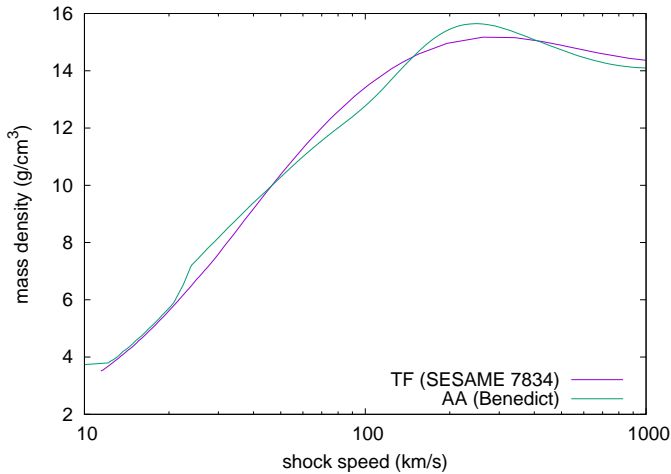}\end{center}
\caption{Principal shock Hugoniot of carbon (initially full-density diamond)
   deduced from previous EOS, in $u_s$-$\rho$ space.}
\label{fig:hugusdtest}
\end{figure}

We considered the principal shock Hugoniot from multiple EOS.
Irrespective of the underlying models and of the actual loci,
$\rho_s(u_s)$ was found to vary smoothly and slowly in the regime
explored by these experiments (Fig.~\ref{fig:hugusdtest}). 
We therefore used low-order polynomials 
to represent $\rho_s(u_s)$ in parameterized form.

\begin{figure}
\begin{center}\includegraphics[scale=0.72]{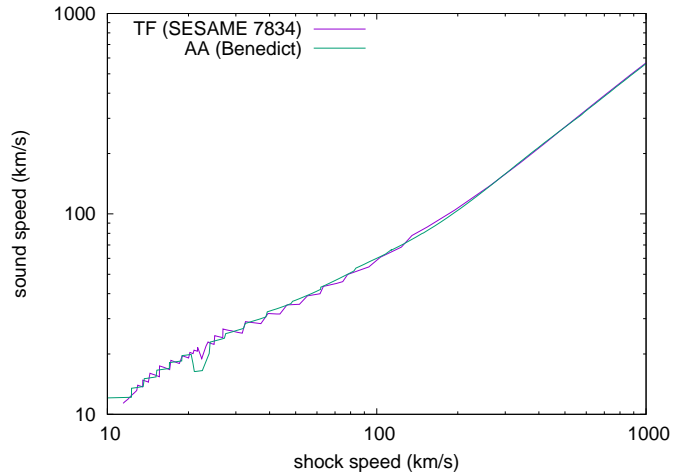}\end{center}
\caption{Sound speed along the principal shock Hugoniot of carbon (initially full-density diamond)
   deduced from previous EOS.}
\label{fig:hugusctest}
\end{figure}

Similarly,
for the density variation behind the shock, Eq.~\ref{eq:acc},
we considered the variation of sound speed with shock speed,
$c(u_s)$, from example EOS.
In the regime of these experiments, $c(u_s)$ was predicted to be
almost indistinguishable from a straight line (Fig.~\ref{fig:hugusctest}), 
and thus low-order polynomials also
appear suitable for representing this Hugoniot function in parameterized form.

We note that, even if any given functional forms for $\rho(u_s)$ and
$c(u_s)$ are unsuitable for describing these parameters over the full Hugoniot,
they are always able to represent a sufficiently small portion of the Hugoniot
as a Taylor expansion about any arbitrary pressure.
This result is hydrodynamically rigorous, i.e. it does not introduce
any approximations.

If shock experiments are affected by non-hydrodynamic effects such as preheat
from the hohlraum, the deduced $\rho(u_s)$ relation would not be unique.
Thus, preheat may be identified 
by comparing experiments with different drive pressures and thus
different levels of preheat at a given location along the Hugoniot,
such as from the low and high drive shots considered here.

\subsection{Parameterize EOS}
In moving from an explicitly density field reconstruction $\rho(r,t)$ to
a reconstruction via Hugoniot functions $\rho(u_s)$ and $c(u_s)$,
we are replacing an experiment-specific model with a model defined via
functions related to the EOS.
A further step is to define the model directly via analytic EOS functions.
We investigated this approach with Gr\"uneisen EOS defined in a way
commonly used in the past for data obtained from shock experiments on solids
up to a few hundred gigapascals \cite{Steinberg1996},
through the relation between shock and particle speeds $u_s(u_p)$,
and the Gr\"uneisen parameter expressed as a function of mass density
$\Gamma(\rho)$,
\begin{equation}
p(\rho,e)=p_r(\rho)+\Gamma(\rho)\left[e-e_r(\rho)\right]
\end{equation}
where $p_r(\rho)$ and $e_r(\rho)$ constitute a reference locus, in this
case the Hugoniot inferred from $u_s(u_p)$ and the Rankine-Hugoniot relations.
We implemented this approach as a further step in the Hugoniot function method
above.
The Rankine-Hugoniot relations were solved to deduce $\rho(u_s)$ from
$u_s(u_p)$, and the derivatives of the Gr\"uneisen EOS were used to
calculate $c(u_s)$. 
When this procedure is valid, it has the advantage of producing EOS that 
may be used directly in many hydrocodes, which often have versions of the
Gr\"uneisen EOS based on a polynomial $u_s(u_p)$ relation and variants of
a power law form for $\Gamma(\rho)$.

\begin{figure}
\begin{center}\includegraphics[scale=0.72]{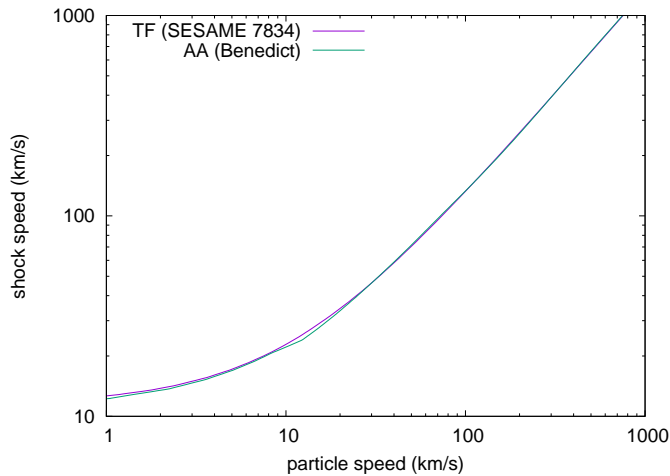}\end{center}
\caption{Relation between shock and particle speeds
   along the principal shock Hugoniot of carbon (initially full-density diamond)
   deduced from previous EOS.}
\label{fig:usuptest}
\end{figure}

\begin{figure}
\begin{center}\includegraphics[scale=0.72]{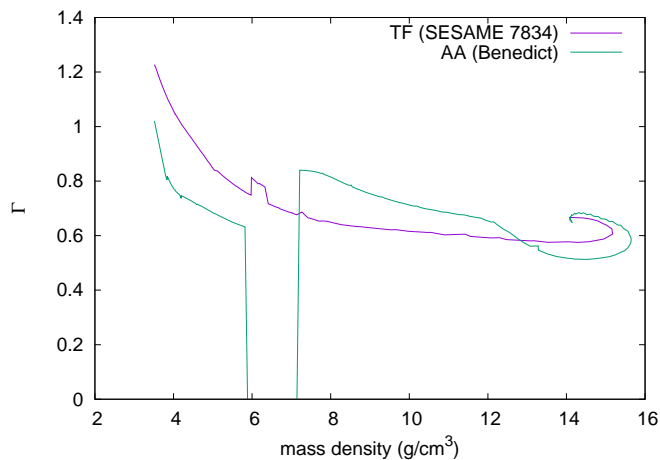}\end{center}
\caption{Gr\"uneisen parameter
   along the principal shock Hugoniot of carbon (initially full-density diamond)
   deduced from previous EOS.}
\label{fig:gammarhotest}
\end{figure}

As for the Hugoniot functions, we investigated the behavior of these
functions using EOS constructed with different physical assumptions.
In the regime of our experiments, the $u_s(u_p)$ relation was found to be
very close to a straight line (Fig.~\ref{fig:usuptest}). 
The Gr\"uneisen parameter $\Gamma$ was found to
have a more complicated variation than found for the other functions
(Fig.~\ref{fig:gammarhotest}).
In particular, it became multivalued at densities around peak compression,
where the simplification that $\Gamma$ is a function of $\rho$ only 
(and not temperature, or energy deviation from the reference) breaks down.
In any case, any EOS using the principal Hugoniot as a reference must break
down by peak compression as the Hugoniot cannot be used as a reference
for states of higher density than peak compression.

As with the Hugoniot function approach,
even if the Gr\"uneisen EOS cannot capture the wide-range structure of
the Hugoniot, it still represents a Taylor expansion 
about a state on the Hugoniot, and can thus be used in a local sense, 
for instance by splitting the radiograph into multiple sections.
Similarly, the use of an EOS imposes uniqueness on the deduced Hugoniot,
which may be incorrect if different experiments are affected by different
levels of preheat, but this can again be assessed by the quality of fit to
the radiographs.

\section{Analysis of carbon experiments}
The outer surface of the diamond was evident in both experiments,
so the opacity could be inferred simultaneously with the density.
The streak radiograph was analyzed to deduce
the radius-time distribution of mass density,
represented as discussed above.
The locus of the shock was represented by the function
\begin{equation}
r_s(t)=\alpha(t_c-t)^\beta
\end{equation}
where $\alpha$, $\beta$, and $t_c$ were fitting parameters.
The locus of the marker layer was represented by the function
\begin{equation}
r_m(t)=r_{\mbox{min}}+\alpha e^{-\beta t}
\end{equation}
where $\alpha$, $\beta$, and $r_{\mbox{min}}$ were fitting parameters.
Quadratics were found adequate to represent $\rho(u_s)$,
and straight lines for $c(u_s)$. 
The locus and value of the maximum density, and the value at the
outer surface, were represented as before.
These functions were able to capture these loci over the full range of
the record.

\begin{figure}
\begin{center}\includegraphics[width=0.5\textwidth]{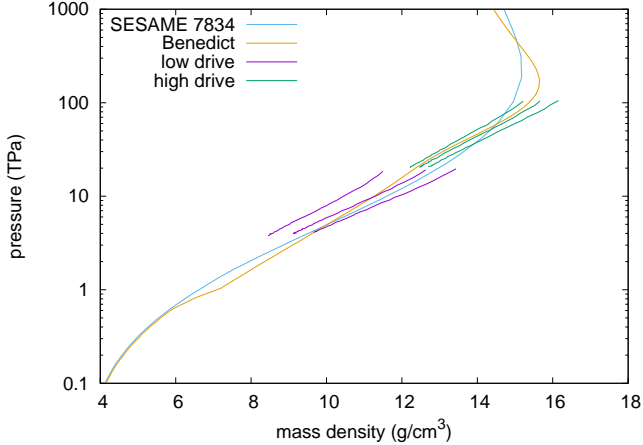}\end{center}
\caption{Hugoniot for carbon (initially diamond) deduced using
   the density field model.
   Thin lines are $1\sigma$ contours for the corresponding thick line.}
\label{fig:hugdp1}
\end{figure}

\begin{figure}
\begin{center}\includegraphics[width=0.5\textwidth]{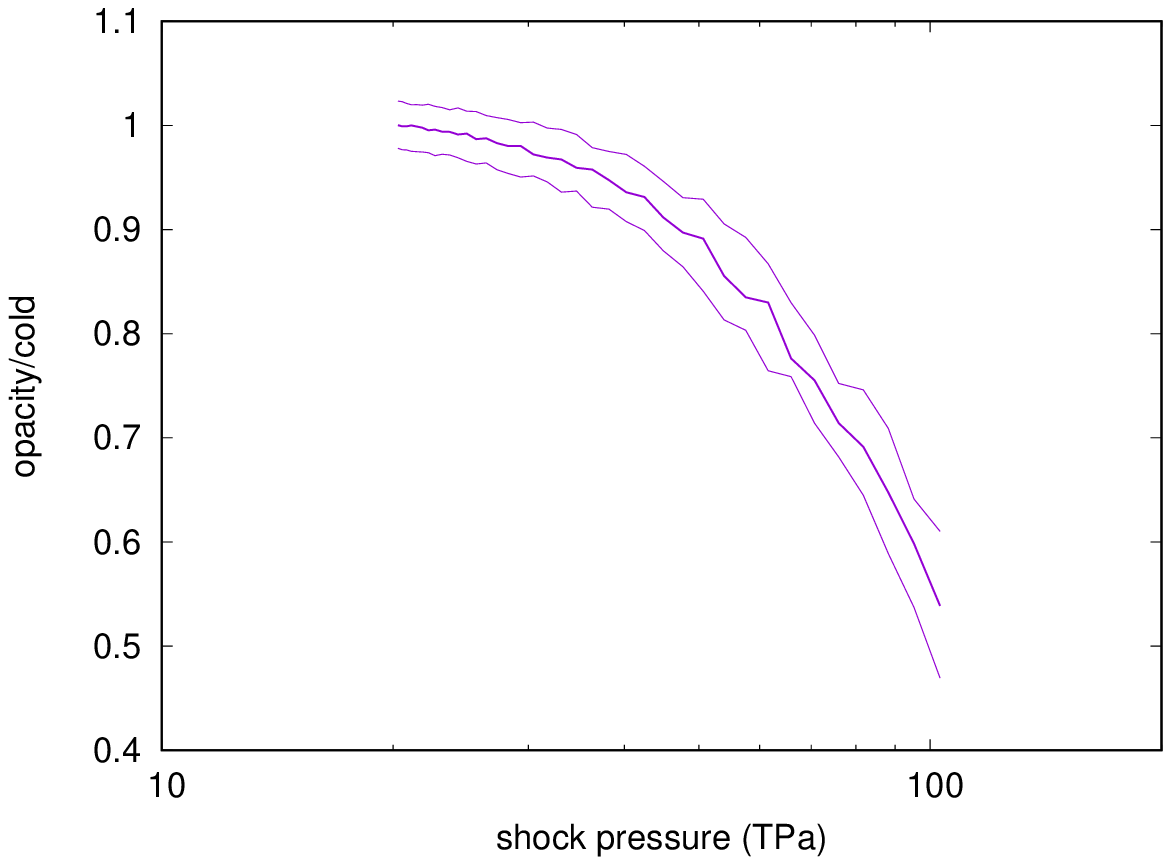}\end{center}
\caption{Opacity of carbon (initially diamond) to 9\,keV x-rays,
   along the principal Hugoniot using the density field model.
   Thin lines are $1\sigma$ contours for the corresponding thick line.}
\label{fig:hugpmu1}
\end{figure}

\begin{figure}
\begin{center}\includegraphics[width=0.5\textwidth]{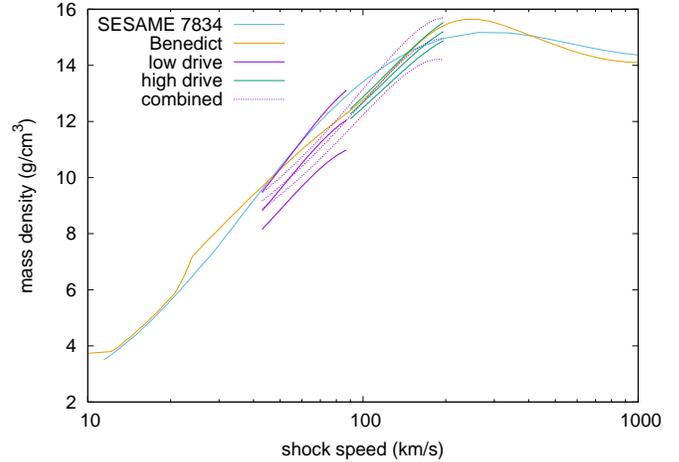}\end{center}
\caption{Hugoniot function $\rho(u_s)$ fit for carbon (initially diamond).
   Thin lines are $1\sigma$ contours for the corresponding thick line.}
\label{fig:hugusd2}
\end{figure}

\begin{figure}
\begin{center}\includegraphics[width=0.5\textwidth]{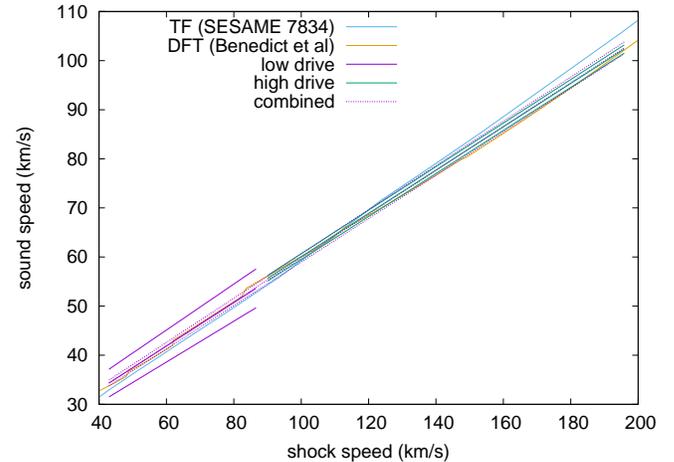}\end{center}
\caption{Fit for sound speed along the Hugoniot for carbon (initially diamond).
   Thin lines are $1\sigma$ contours for the corresponding thick line.}
\label{fig:hugusc2}
\end{figure}

\begin{figure}
\begin{center}\includegraphics[width=0.5\textwidth]{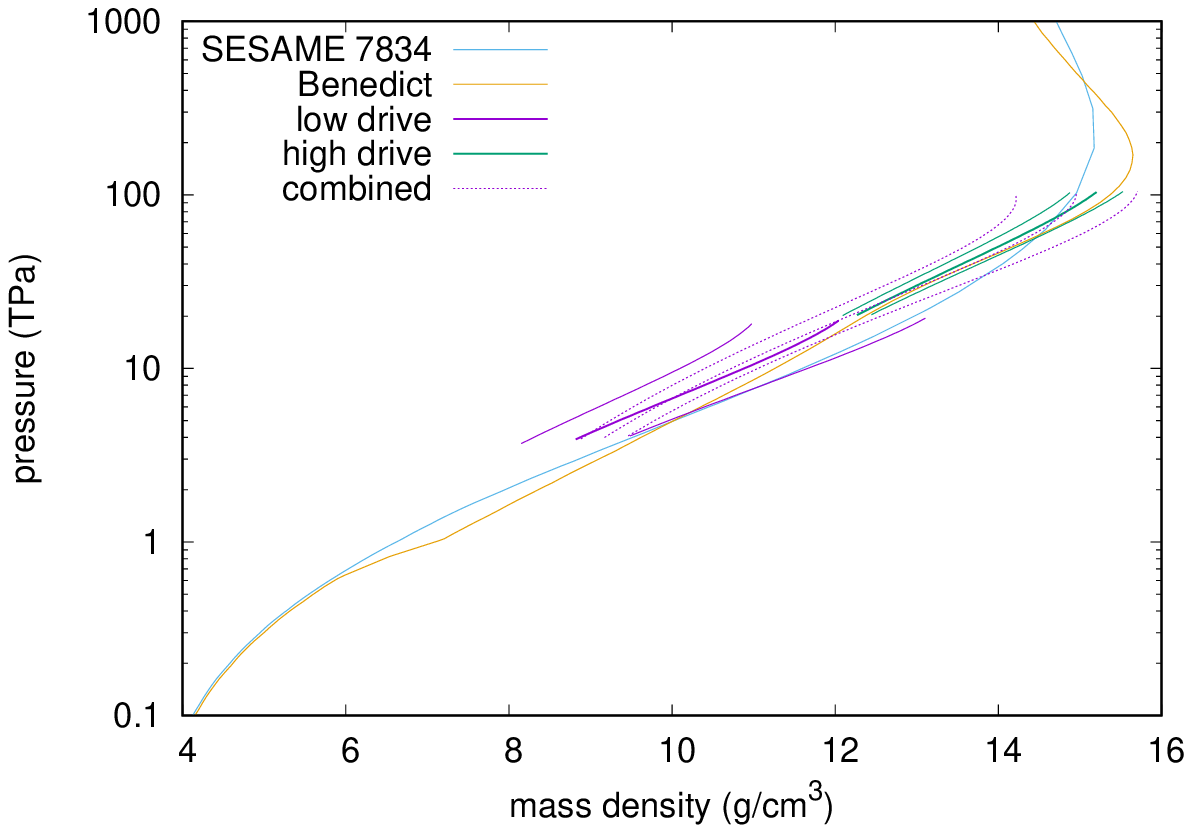}\end{center}
\caption{Hugoniot for carbon (initially diamond) deduced using
   the Hugoniot function model.
   Thin lines are $1\sigma$ contours for the corresponding thick line.}
\label{fig:hugdp2}
\end{figure}

\begin{figure}
\begin{center}\includegraphics[width=0.5\textwidth]{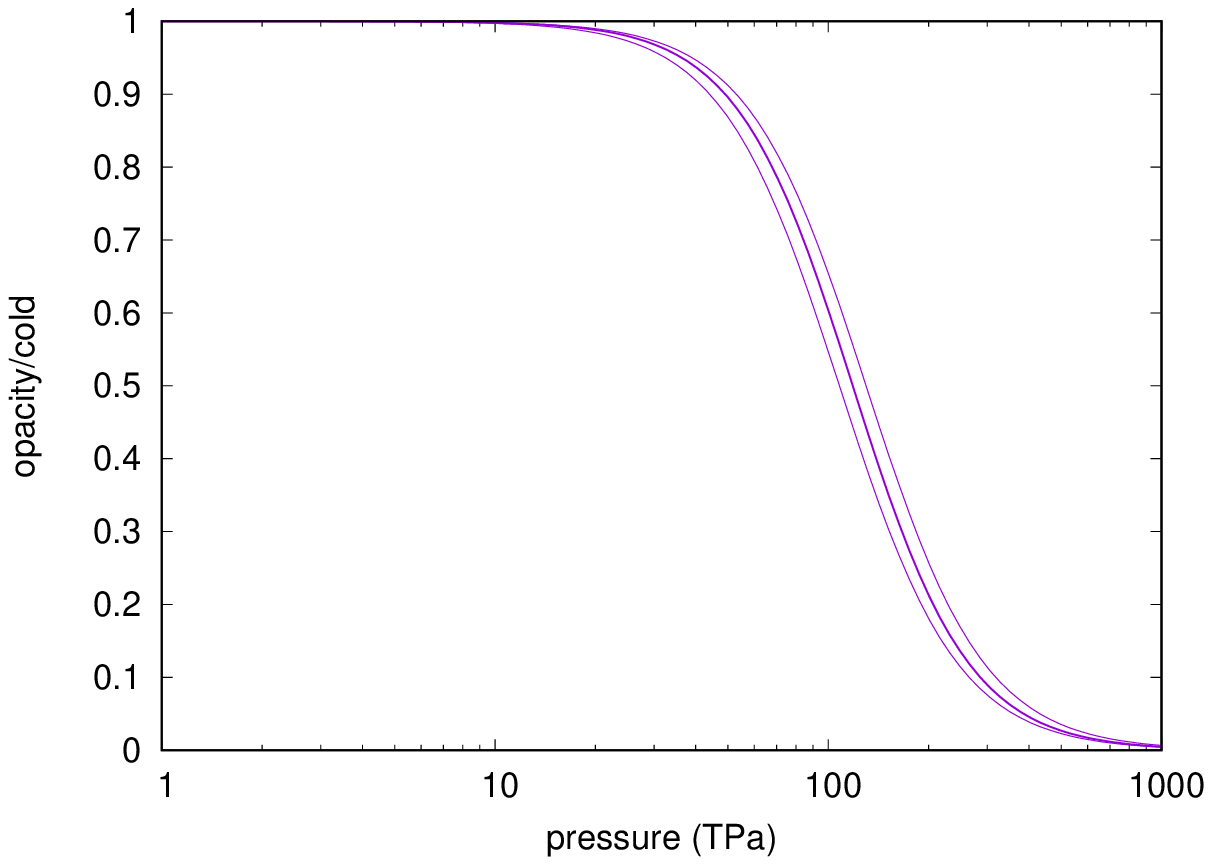}\end{center}
\caption{Opacity of carbon (initially diamond) to 9\,keV x-rays,
   along the principal Hugoniot using the Hugoniot function model.
   Thin lines are $1\sigma$ contours for the corresponding thick line.}
\label{fig:hugpmu2}
\end{figure}

\begin{figure}
\begin{center}\includegraphics[width=0.5\textwidth]{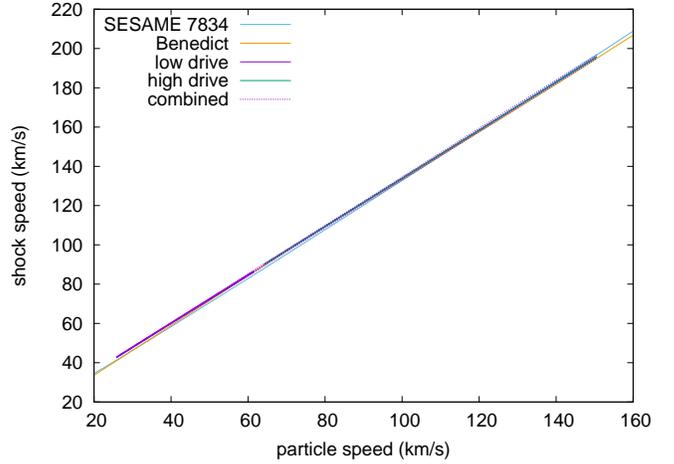}\end{center}
\caption{Shock speed -- particle speed fit for carbon (initially diamond).
   Thin lines are $1\sigma$ contours for the corresponding thick line.}
\label{fig:hugupus3}
\end{figure}

\begin{figure}
\begin{center}\includegraphics[width=0.5\textwidth]{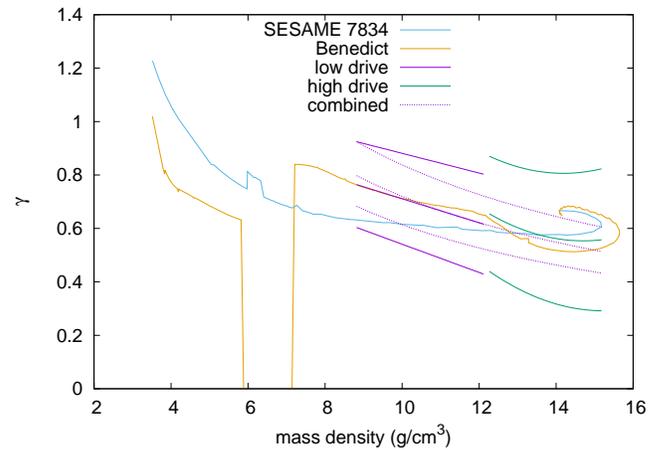}\end{center}
\caption{Fit for Gr\"uneisen parameter along the Hugoniot for carbon (initially diamond).
   Thin lines are $1\sigma$ contours for the corresponding thick line.}
\label{fig:hugdgam3}
\end{figure}

\begin{figure}
\begin{center}\includegraphics[width=0.5\textwidth]{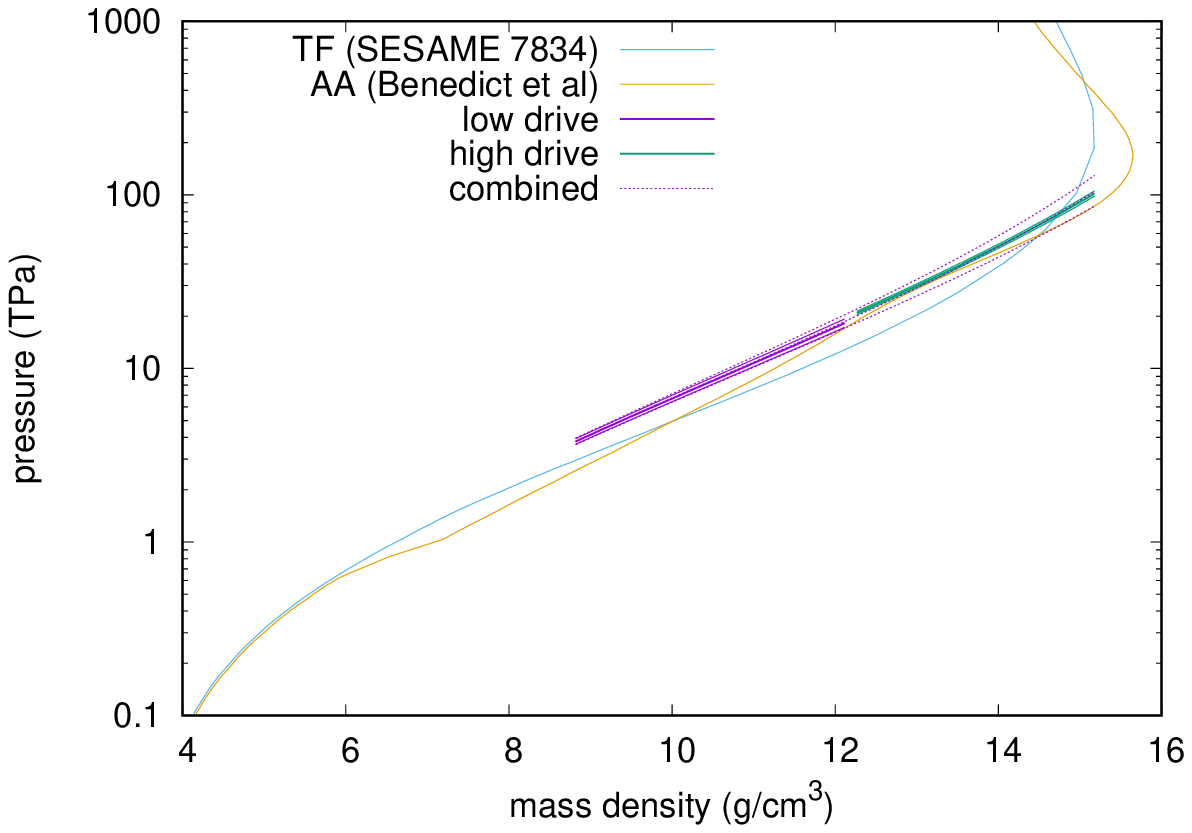}\end{center}
\caption{Hugoniot for carbon (initially diamond) deduced using
   the EOS function model.
   Thin lines are $1\sigma$ contours for the corresponding thick line.}
\label{fig:hugdp3}
\end{figure}

\begin{figure}
\begin{center}\includegraphics[width=0.5\textwidth]{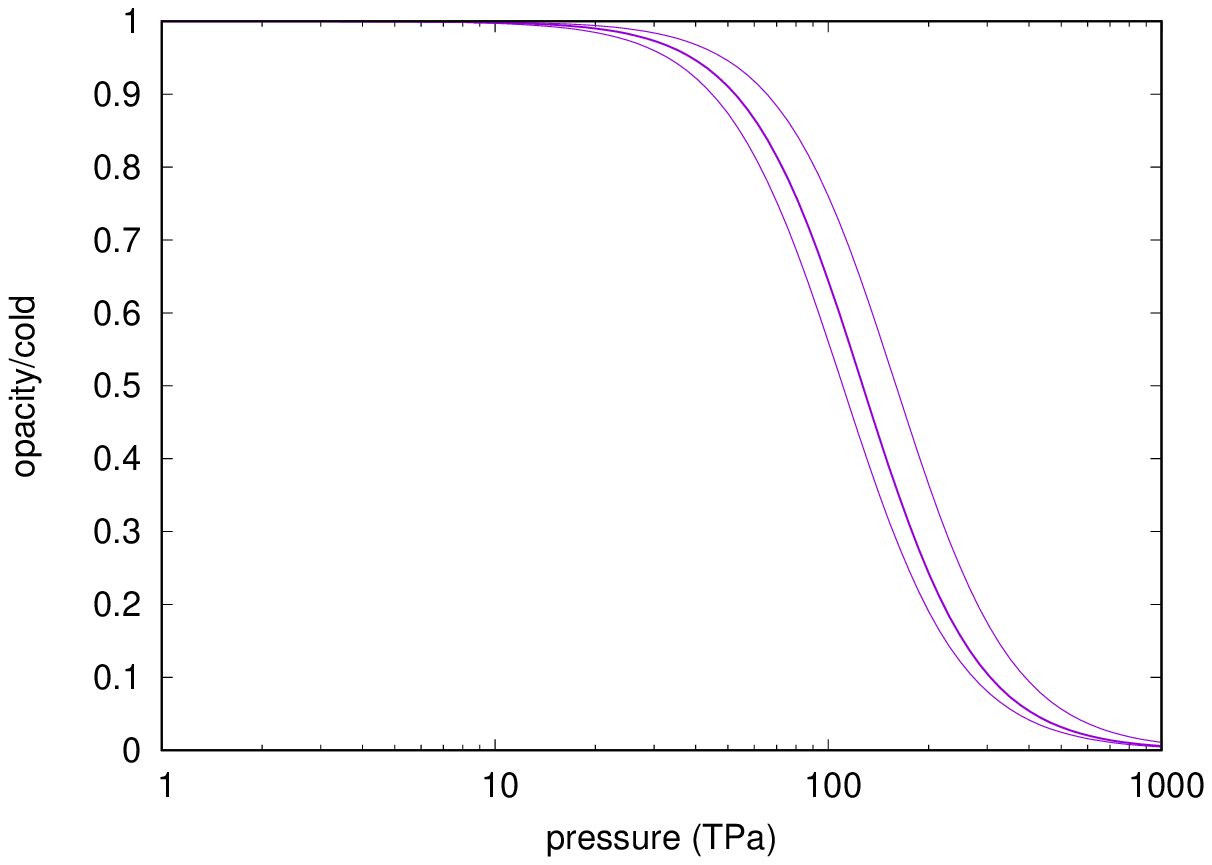}\end{center}
\caption{Opacity of carbon (initially diamond) to 9\,keV x-rays,
   along the principal Hugoniot using the EOS function model.
   Thin lines are $1\sigma$ contours for the corresponding thick line.}
\label{fig:hugpmu3}
\end{figure}

Analyses were performed for each shot separately, and also for both
shots together.
The opacity from shot N161016-3 was consistent with it remaining cold,
and so the Hugoniot states from this shot were determined assuming the
cold opacity, as this gave a smaller uncertainty.
The analysis was performed using the density field reconstruction
(Figs~\ref{fig:hugdp1} and \ref{fig:hugpmu1}),
the Hugoniot functions
(Figs~\ref{fig:hugusd2} to \ref{fig:hugpmu2}),
and the EOS functions
(Figs~\ref{fig:hugupus3} to \ref{fig:hugpmu3}).
None of the $\Gamma(\rho)$ functions tried was a good fit to
the full range of the data: a linear variation gave the best match to the
low drive shot; a quadratic variation to the high drive, and a power law 
to the simultaneous fit to both shots.
Using the present method of deducing the best fit and uncertainties,
the nominal uncertainty in the deduced shock Hugoniot was lowest for the 
Gr\"uneisen EOS fit.
However, the absolute goodness of fit along the best solution was significantly
lower than for Hugoniot function fit, meaning that the best-fitting residual
difference between the simulated and measured radiographs was larger.
In other words, the Gr\"uneisen EOS was less able to fit the overall flow
distribution probed by the radiograph.
This deficiency seems to be related to the use of a Gr\"uneisen parameter
varying with $\rho$ only, as the deduced $u_s-u_p$ was indistinguishable from
linear.
The overall goodness of fit from the nominal best Gr\"uneisen EOS fit
was consistent with the Hugoniot function fit in the same area of the $p-\rho$
plane.
The main difference between the two analysis approaches is in the use of
a $c(u_s)$ or $\Gamma(\rho)$ relation.
Considering the theoretical EOS models above, the $\Gamma(\rho)$ relation
is inappropriate around peak Hugoniot compression, where it varies rapidly
and is double-valued, whereas the $c(u_s)$ relation is smooth and monotonic.
However, the main difference in deduced Hugoniots was for the low drive shot,
which is surprising.
Shot N161016-3 was unusually noisy, and possibly the Gr\"uneisen EOS model
turns out to be more sensitive to noise because it required more
parameters to describe $\Gamma(\rho)$ than the Hugoniot function required
to describe $c(u_s)$.

\section{Conclusions}
By modifying the structure of the model used to describe the distribution
of mass density when reconstructing time-resoilved radiographs of converging
shocks, so that the parameters were in functions relating to the EOS
rather than experiment-specific spatial and temporal functions,
we were able to significantly increase the precision and amount 
of EOS data deduced from these experiments while also improving the 
numerical stability and efficiency of the analysis, without loss of generality
in the reconstruction of density from a given experiment.
This approach also enabled the shock Hugoniot to be inferred using 
data from multiple experiments.
The precision was increased further through the use of a parameterized
function for the x-ray opacity, instead of the unfolding reconstruction
used previously.

For carbon (initially diamond), the nominal uncertainty 
in the deduced Hugoniot was lower when analyzed using
an analytic Gr\"uneisen EOS with the Hugoniot as reference
rather than
when using functions for $\rho(u_s)$ and $c(u_s)$.
The opposite was true for the opacity.
Both techniques had significantly lower uncertainty than the previous
density field model and an opacity deduced directly from the change in
apparent mass.
However, the analysis using the Hugoniot functions $\rho(u_s)$ and $c(u_s)$
gave a significantly better goodness of fit than the analytic Gr\"uneisen EOS,
which may be because a Gr\"uneisen parameter $\Gamma$ varying with
$\rho$ only was insufficient to represent the EOS of carbon over the range
of states explored in even the lower-drive experiment.

Hugoniot data for carbon were obtained to $\sim$80\,TPa, by which point the
opacity had fallen to a level implying a $k$-shell ionization of around 30\%.
The uncertainty was closer to what had originally been predicted 
for these experiments
from an assessment of the expected sources of error.

\section*{Acknowledgments}
This work was performed under the auspices of
the U.S. Department of Energy under contract DE-AC52-07NA27344.


\begin{thebibliography}{10}
\bibitem{shock}{For example,
   R.G.~McQueen et al, 
   in R.~Kinslow (Ed.),
   ``High Velocity Impact Phenomena''
   (Academic Press, New York, 1970).}
\bibitem{Kritcher2014}{A.L.~Kritcher, T.~D\"oppner, D.~Swift, J.~Hawreliak, G.~Collins, J.~Nilsen, B.~Bachmann,E.~Dewald, D.~Strozzi, S.~Felker, O.L.~Landen, O.~Jones, C.~Thomas, J.~Hammer, C.~Keane, H.J.~Lee, S.H.~Glenzer, S.~Rothman, D.~Chapman, D.~Kraus, P.~Neumayer, and R.W.~Falcone,
   High Energy Density Phys. {\bf 10}, pp~27--34 (2014).}
\bibitem{Doeppner2018}{T.~D\"oppner, D.C.~Swift, A.L.~Kritcher, B.~Bachmann,
   G.W.~Collins, D.A.~Chapman, J.~Hawreliak, D.~Kraus, J.~Nilsen, S.~Rothman,
   L.X.~Benedict, E.~Dewald, D.E.~Fratanduono, J.A.~Gaffney, S.H.~Glenzer,
   S.~Hamel, O.L.~Landen, H.J.~Lee, S.~LePape, T.~Ma, M.J.~MacDonald,
   A.G.~MacPhee, D.~Milathianaki, M.~Millot, P.~Neumayer, P.A.~Sterne,
   R.~Tommasini, and R.W.~Falcone,
   Phys. Rev. Lett. {\bf 121}, 025001 (2018).}
\bibitem{Swift2018}{D.C.~Swift, A.L ~Kritcher, J.A ~Hawreliak, A.~Lazicki,
   A.~MacPhee, B.~Bachmann, T.~D\"oppner, J.~Nilsen, G.W ~Collins, S.~Glenzer,
   S.D ~Rothman, D.~Kraus, and R.W ~Falcone,
   Rev. Sci. Instrum. {\bf 89}, 053505 (2018).}
\bibitem{Kritcher2020}{A.L.~Kritcher, D.C.~Swift, T.~D\"oppner, B.~Bachmann, L.X.~Benedict, G.W.~Collins, J.L.~DuBois, F.~Elsner, G.~Fontaine, J.A.~Gaffney, S.~Hamel, A.~Lazicki, W.R.~Johnson, N.~Kostinski, D.~Kraus, M.J.~MacDonald, B.~Maddox, M.E.~Martin, P.~Neumayer, A.~Nikroo, J.~Nilsen, B.A.~Remington, D.~Saumon, P.A.~Sterne, W.~Sweet, A.A.~Correa, H.D.~Whitley, R.W.~Falcone, and S.H.~Glenzer,
   Nature {\bf 584}, 7819, pp~51-54 (2020).}
\bibitem{Swift2021}{D.C.~Swift, A.L.~Kritcher, J.A.~Hawreliak, J.~Gaffney, A.~Lazicki, A.~MacPhee, B.~Bachmann, T.~Doeppner, J.~Nilsen, H.D.~Whitley, G.W.~Collins, S.~Glenzer, S.D.~Rothman, D.~Kraus, and R.W.~Falcone,
   Rev. Sci. Instrum. {\bf 92}, 063514 (2021).}
\bibitem{hohlraum}{S.W.~Haan, P.A.~Amendt, T.R.~Dittrich, B.A.~Hammel, S.P.~Hatchett, M.C.~Herrmann, O.A.~Hurricane, O.S.~Jones, J.D.~Lindl, M.M.~Marinak, D.~Munro, S.M.~Pollaine, J.D.~Salmonson, G.L.~Strobel, and L.J.~Suter,
   Nucl.~Fusion {\bf 44}, S171 (2004).}
\bibitem{icf}{J.~Lindl,
   Phys. Plasmas 2, 3933 (1995).}
\bibitem{Hopkins2015}{L.F.~Berzak Hopkins, N.B.~Meezan, S.~Le Pape, L.~Divol, A.J.~Mackinnon, D.D.~Ho, M.~Hohenberger, O.S.~Jones, G.~Kyrala, J.L.~Milovich, A.~Pak, J.E.~Ralph, J.S.~Ross, L.R.~Benedetti, J.~Biener, R.~Bionta, E.~Bond, D.~Bradley, J.~Caggiano, D.~Callahan, C.~Cerjan, J.~Church, D.~Clark, T.~D\"oppner, R.~Dylla-Spears, M.~Eckart, D.~Edgell, J.~Field, D.N.~Fittinghoff, M.~Gatu Johnson, G.~Grim, N.~Guler, S.~Haan, A.~Hamza, E.P.~Hartouni, R.~Hatarik, H.W.~Herrmann, D.~Hinkel, D.~Hoover, H.~Huang, N.~Izumi, S.~Khan, B.~Kozioziemski, J.~Kroll, T.~Ma, A.~MacPhee, J.~McNaney, F.~Merrill, J.~Moody, A.~Nikroo, P.~Patel, H.F.~Robey, J.R.~Rygg, J.~Sater, D.~Sayre, M.~Schneider, S.~Sepke, M.~Stadermann, W.~Stoeffl, C.~Thomas, R.P.J.~Town, P.L.~Volegov, C.~Wild, C.~Wilde, E.~Woerner, C.~Yeamans, B.~Yoxall, J.~Kilkenny, O.L.~Landen, W.~Hsing, and M.J.~Edwards,
   Phys. Rev. Lett. {\bf 114}, 175001 (2015).}
\bibitem{Dewald2004}{E.L.~Dewald, K.M.~Campbell, R.E.~Turner, J.P.~Holder, O.L.~Landen, S.H.~Glenzer, R.L.~Kauffman, L.J.~Suter, M.~Landon, M.~Rhodes, and D.~Lee,
   Rev. Sci. Instrum. {\bf 75}, 3759 (2004).}
\bibitem{xbt}{E.L.~Dewald et al, 
   Phys. Rev. Lett {\bf 111}, 235001 (2013).}
\bibitem{hydra}{M.M.~Marinak, S.W.~Haan, T.R.~Dittrich, R.E.~Tipton, and G.B.~Zimmerman,
   Phys. Plasmas {\bf 5}, 1125 (1998).}
\bibitem{Abel1826}{N.H.~Abel, J.~reine \&\ angewandte Math., {\bf 1},
   pp.~153-157 (1826).}
\bibitem{Cunningham1995}{G. Cunningham, K. Hanson, G. Jennings, and D. Wolf,
   Rev. Prog. Quant. {\bf NDE 14A}, 747 (1995).}
\bibitem{dac}{For example, A.D.~Chijioke,
   W.J.~Nellis, A.~Soldatov, and I.F.~Silvera,
   {\it The ruby pressure standard to 150 GPa},
   J.~Appl. Phys. {\bf 98}, 114905 (2005).}
\bibitem{ariane}{Jay Ayers et al,
   Proc. SPIE. {\bf 8505},
   85050J (2012).}
\bibitem{MacPhee2014}{A.~MacPhee, unpublished (2014).}
\bibitem{Fox1992}{For instance, R.W.~Fox and A.T.~MacDonald,
   {\it Introduction to Fluid Mechanics}
   (Wiley, Hoboken NJ, 1992).}
\bibitem{Benson1992}{D.~Benson,
   Computer Methods in Appl. Mechanics and Eng. {\bf 99}, 235 (1992).}
\bibitem{ses7592}{J.~Barnes and S.~Lyon, documentation for {\sc Sesame}
   EOS 7592, unpublished (1988).}
\bibitem{opal}{C.A.~Iglesias, Astrophys. J. {\bf 464}, 943 (1996).}
\bibitem{imp}{S.J. Rose, J. Phys. B: At. Mol. Opt. Phys. {\bf 25}, 1667 (1992).}
\bibitem{Marsh1980}{S.P.~Marsh (Ed), {\it LASL Shock Hugoniot Data}
   (University of California, Berkeley, 1980).}
\bibitem{Steinberg1996}{D.J.~Steinberg,
   {\it Equation of state and strength parameters for selected materials},
   Lawrence Livermore National Laboratory report UCRL-MA-106439 change 1
   (1996).}
\end{thebibliography}
\end{document}